\documentclass[]{article}
\usepackage[]{emulateapj}
\usepackage[]{times}

\newenvironment{tablehere}
  {\def\@captype{table}}
  {}
\makeatother

\begin{document}

\title{Frequencies of $f$ modes in differentially
rotating relativistic stars and secular stability limits}

\author{Shin'ichirou Yoshida,  Luciano Rezzolla\altaffilmark{1}}
\affil{SISSA, International School for Advanced Studies, Via Beirut 2-4,
	34014 Trieste, Italy}

\and 

\author{Shigeyuki Karino,  Yoshiharu Eriguchi}
\affil{Department of Earth Science and Astronomy,
Graduate School of Arts and Sciences, University of
Tokyo, Komaba, Tokyo 153-8902, Japan}

\altaffiltext{1}{INFN, Department of Physics, University
        of Trieste, Via Valerio, 2 34127 Trieste, Italy}

\date{\today}

\begin{abstract}
We have computed the eigenfrequencies of $f$ modes
for a constant-rest-mass sequences of rapidly rotating
relativistic inviscid stars in differential rotation. The
frequencies have been calculated neglecting the metric
perturbations (the relativistic Cowling approximation) and
expressed as a function of the ratio between the
rotational kinetic energy and the absolute value of the
gravitational energy of the stellar model $\beta \equiv
T/|W|$. The zeros and the end-points of these sequences
mark respectively the onset of the secular instability
driven by gravitational radiation-reaction and the
maximum value of $\beta$ at which an equilibrium model
exists. In differentially rotating stars the secular
stability limits appear at a $\beta$ larger than those
found for uniformly rotating stars. Differential
rotation, on the other hand, also allows for the
existence of equilibrium models at values of $\beta$
larger than those for uniformly rotating stars, moving
the end-point of the sequences to larger $\beta$. As a
result, for some degrees of differential rotation, the
onset of the secular instability for $f$ modes is
generally favoured by the presence of differential
rotation.
\end{abstract}

\keywords{stars:rotation --- stars:oscillation --- stars:neutron}


\section{Introduction}

	Instabilities of rotating relativistic stars have
been studied for more than three decades in relation to
the emission of gravitational waves driven by
radiation-reaction. Since the discovery of the so called
Chandrasekhar-Friedman-Schutz (CFS)
instability (Chandrasekhar 1970 ; Friedman and Schutz 1978), 
in fact, the stability
properties of a number of non-axisymmetric stellar
oscillations have been investigated using different
techniques (see Friedman 1998 ; Andersson and Kokkotas
2001 for recent reviews).
The $f$ (fundamental) modes were among the first modes of
oscillations ever to be studied in rotating relativistic
stars. These are spheroidal modes with harmonic indices
$l=m$ and represent the generalization of the Kelvin
modes of Newtonian Maclaurin spheroids to compressible
fluid stars. Over the years,
the literature on the subject has been continuously
updated and the limits for the onset of the instability
have been improved through different approaches. This has
been done by several groups focussing on Newtonian
stellar
models (Bardeen et al. 1977 ; Comins 1979a, 1979b ; Clement
1979 ; Durisen and Imamura 1981 ; Imamura et al. 1985
Managan 1985 ; Ipser and Lindblom 1990)
on post-Newtonian stellar models (Cutler and Lindblom 1992)
and on rapidly rotating relativistic
stars (Yoshida and Eriguchi 1997 ; Stergioulas and Friedman
1998 ; Morsink et al. 1999 ; Yoshida and Eriguchi 1999).
Part of the interest
in these modes comes from the fact that they have long been
regarded as the modes of oscillation most susceptible to
the CFS instability. In particular, for stars rotating at
a rate at which mass starts being shed at the equator
(the mass-shedding limit), general relativistic calculations
on the stability limit indicate that the $l=m=2$, $f$-mode
(also referred to as the ``bar-mode'') could have the
shortest growth timescale. Because of this, and because of a
possible weakening of the bulk viscosity at very high
temperatures (Lai 2001), the bar-mode may represent the
most important non-axisymmetric instability in very hot
and rapidly rotating newly born neutron stars.

	The occurrence of an $f$-mode instability in a
newly born uniformly rotating neutron star is in general
prevented by the large bulk and shear viscosities of
nuclear matter in those conditions and detailed
calculations have shown that the instability is
suppressed except for very large rotation rates, close to
the mass-shedding limit (Ipser and Lindblom 1991); for superfluid
stars an even stronger damping was
calculated (Lindblom and Mendel 1995). More recently, however, a number
of new elements have improved our understanding of the
instability and have again increased the expectations
that the $f$ mode instability might characterize the
earliest life stages of a newly formed neutron star. The
first of these new elements was provided by Cutler and
Lindblom (1992)
who have shown that, within a post-Newtonian
approximation, general relativistic effects tend to
further destabilize the $f$ mode, lowering the critical
value of the ratio between the stellar rotational kinetic
energy and the absolute value of the gravitational
energy, $\beta_c \equiv (T/|W|)_c$ at which the secular
$f$-mode instability is triggered. Analogous results have
also been found by Yoshida and Eriguchi (1997) within the relativistic
Cowling approximation and by Stergioulas and Friedman (1998)
in fully
general relativistic calculations. The second new element
was provided by Shibata and Ury\={u} (2000) whose fully general
relativistic hydrodynamical simulations have shown that
the remnants of binary neutron star mergers could be, at
least for polytropic equations of state, rapidly and
differentially rotating stars. In addition to this,
Liu and Lindblom (2001) have recently computed the structure of
objects formed in accretion-induced collapse of rotating
white dwarfs and found that these objects can rotate
extremely rapidly and differentially. Furthermore, a
differentially rotating object could be produced also
during the process of core-collapse in supernova leading to the
formation of a neutron star.

	Within this framework, differential rotation has
two major consequences. Firstly, it allows for the
existence of an equilibrium model at values of $\beta$ which
are considerably larger than the ones supported by the
counterparts with the same rest-mass but uniform
rotation (Baumgarte et al. 2000). Secondly, as shown by a number of
authors for Newtonian stellar
models (Managan 1985, Imamura et al. 1985, Imamura and Durisen 2001),
it increases the
critical value $\beta_c$ for the onset of the secular
instability. In this revised picture, we have computed
the eigenfrequencies of $f$ modes and determined the
secular stability limits for rapidly rotating
relativistic stars with differential rotation.

\section{Formulation}

	The equilibrium stellar models are assumed to be
stationary and axisymmetric and are constructed with a
numerical code based on the method of 
Komatsu et al. (1989a,b).
Their spacetime is therefore described by
the line element
\begin{equation}
\label{eq_metric}
ds^2 = -e^{2\nu} dt^2 + e^{2\alpha} (dr^2 + r^2d\theta^2)
	+ e^{2\mu}r^2\sin^2\theta (d\phi-\omega dt)^2\ ,
\end{equation}
where $\nu, \alpha, \mu,$ and $\omega$ are the
``gravitational potentials'' and are functions of the $r$
and $\theta$ coordinates only. The stars are modeled as
relativistic polytropes with the equation of state (EOS)
\begin{equation}
\frac{p}{\rho_c} = \kappa 
	\left(\frac{\rho}{\rho_c}\right)^{1+1/N}\ ,	
	\quad \epsilon = \rho + Np \ ,
\end{equation}
where $p, \rho, \epsilon$ are the pressure, the rest-mass
density and the total energy density, respectively. The
subscript ``$c$'' refers to the maximum value for the
equilibrium model which is used for normalization. In
order to investigate how the limits for the secular
instability depend on the ``stiffness'' of the EOS, we
have performed calculations for two different polytropic
models whose properties are summarized in Table~I (We
adopt units in which $c=G=M_\odot=1$.). Along each
sequence of rapidly rotating stellar models, the
polytropic constant $\kappa$, the polytropic index $N$
and the total rest-mass $M_0$ are kept constant.

%
\begin{center}
\begin{tablehere}
\vspace{0.1cm}
\label{tI}
\begin{tabular}{lccc}
\hline
\hline
~~~~~~~~~~~ ~&~ $N$                  ~&~ $\kappa$          
	    ~&~ $M_0/M_\odot$	     \\
\hline
Model $(a)$  ~&~ $0.5$                ~&~ $6.02\times 10^4$
	    ~&~ $1.60$	             \\
Model $(b)$  ~&~ $1.0$                ~&~ $1.00\times 10^2$ 
	    ~&~ $1.52$	             \\
\hline
\end{tabular}
\vspace{0.2cm}
\end{tablehere}
\break
\end{center}%
{\small~~~~TABLE I: Properties of the polytropic equilibrium models.}
\vspace{0.1cm}

	The rapidly rotating stellar models are
constructed after specifying a law of differential
rotation $\Omega = \Omega(r, \theta)$ with a choice which
is, to some extent, arbitrary. This is because there are
a number of differential rotation laws that satisfy the
integrability condition of the equation of hydrostatic
equilibrium as well as the Rayleigh criterion for
dynamical stability (Tassoul 1978). All of these laws are
physically consistent, cannot be excluded on physical
grounds and might influence the qualitative behaviour of
the results. We here follow the formulation suggested by
Komatsu et al. (1989a,b) who have modeled the
rotational angular velocity profile as
\begin{equation}
\label{omegagr}
A_{_{\rm R}}^2(\Omega_0-\Omega) =
	\frac{(\Omega-\omega)r^2\sin^2\theta
	e^{2(\mu-\nu)}}{1-(\Omega-\omega)r^2\sin^2\theta
	e^{2(\mu-\nu)}} \ .
\end{equation}
Here $\Omega_0$ is the angular velocity at the centre of
the coordinate system, i.e. at the intersection of the
rotation axis with the equatorial plane of the
equilibrium models, and $A_{_{\rm R}}$ is a dimensionless
parameter accounting for the degree of differential
rotation. In particular, the degree of differential
rotation increases with $A_{_{\rm R}}^{-1}$, and in the limit of
$A_{_{\rm R}}^{-1} \to 0$, the profile reduces to that of uniform
rotation. In the Newtonian limit, the differential
rotation law (\ref{omegagr}) reduces to the so-called
``j-constant'' law (Clement 1978)
\begin{equation}
\frac{\Omega}{\Omega_0} =
	\frac{A_{_{\rm N}}^2}{A_{_{\rm N}}^2+r^2\sin^2\theta} \ ,
\label{omeganewton}
\end{equation}
and is commonly used in Newtonian calculations
(Eriguchi and M\"{u}ller 1985 ; Karino et al. 2001 ; Rezzolla and Yoshida 2001).

	Once the rapidly and differentially rotating
equilibrium stellar model has been constructed, we
introduce ``adiabatic'' perturbations in the fluid
variables so that the adiabatic index of the perturbed
matter coincides with the polytropic exponent of the
equilibrium configuration
\begin{equation}
\frac{\Delta p}{p} =
	\left(1+\frac{1}{N}\right)\frac{\Delta\rho}{\rho}
	\ .
\end{equation}
Here, $\Delta$ refers to a Lagrangian perturbation with
displacement vector ${\vec \xi}$ and is related to the
Eulerian perturbation $\delta$ of a generic quantity
$\Phi$ through the relation: $\Delta \Phi = \delta \Phi +
{\cal L}_{\vec \xi} \Phi$, with ${\cal L}_{\vec \xi}$
being the Lie derivative along ${\vec
\xi}$ (Friedman and Schutz 1978a).  Because of the stationary and
axisymmetric background, the Eulerian perturbations are
naturally decomposed into a harmonic component of the type
$\sim \exp(-i\sigma t + im\phi)$, where $\sigma$ is the
mode angular frequency and $m$ is an integer. Working
within the Cowling approximation (McDermott et al. 1983), we do
not include Eulerian perturbations in the metric and
the fluid oscillations are investigated in the
fixed background spacetime of the equilibrium
star~(\ref{eq_metric})\footnote{An alternative approach
to the relativistic Cowling approximation was formulated
by Finn (1988)}.  This simplification clearly introduces
an error which is particularly large in the case of the lower
order $f$ modes. However, as discussed in previous works~
(Lindblom and Splinter 1990 ; Yoshida and Kojima 1990 ;
Yoshida and Eriguchi 1997, 1999), the results obtained with
this approximation reproduce well the qualitative
behaviour of the mode eigenfrequencies with the
errors being of the order of 10\% for low mode-numbers
and progressively less for high
mode-numbers (Lindblom and Splinter 1990 ; Yoshida and Eriguchi 1997, 1999).
  Furthermore, by
neglecting the contributions coming from the metric
perturbations, the Cowling approximation tends, at least
for the lowest mode-numbers (i.e. $m=2,3$), to
overestimate the stability (see the comparison
in Figs.~2 and 3).

	Introducing the harmonic perturbations in the
hydrodynamical equations yields a system of four partial
differential equations accounting for baryon number
conservation and conservation of the stress-energy
tensor. These equations are then solved for the four
unknown functions represented by the components of the
Eulerian 3-velocity perturbation 
$\delta v^r, \delta v^\theta, \delta v^\phi$
and the dimensionless quantity $q\equiv\delta
p/(\epsilon+p)$. The partial differential equations are
discretized on a two-dimensional numerical grid and are
solved following the strategy discussed by Yoshida and Eriguchi (1999),
after imposing $\Delta p=0$ at the surface of the
unperturbed star as a boundary condition. For each degree
of differential rotation $A_{_{\rm R}}$ and for each
mode-number $m$, the solution to the eigenvalue problem
is found for increasing values of $\beta$ until the limit
of mass-shedding is found, which is indicated with
$\beta_s$ (see Komatsu et al. 1989a for a definition of
$\beta$).

\section{Results}

	The results of these calculations are presented
in Fig.\ref{fig1}, where we plot the frequencies of the
$m=2$ mode as a function of the parameter $\beta$ for the
model $(b)$ of Table~I. Different curves refer to
different degrees of differential rotation. Note that the
values of $\beta$ at which $\sigma=0$ (i.e. $\beta_c$)
signal the onset of the secular instability (neutral
points) and that these increase as the degree of
differential rotation is increased. At the same time,
however, the differentially rotating models are able to
support larger values of $\beta$ before reaching the mass
shedding limit $\beta_s$, represented by the end-points
of the curves in Fig.\ref{fig1}.

%
%
\vspace{0.2cm} \vbox{ \vskip 0.30truecm
\centerline{\epsfxsize=7.5truecm
\epsfbox{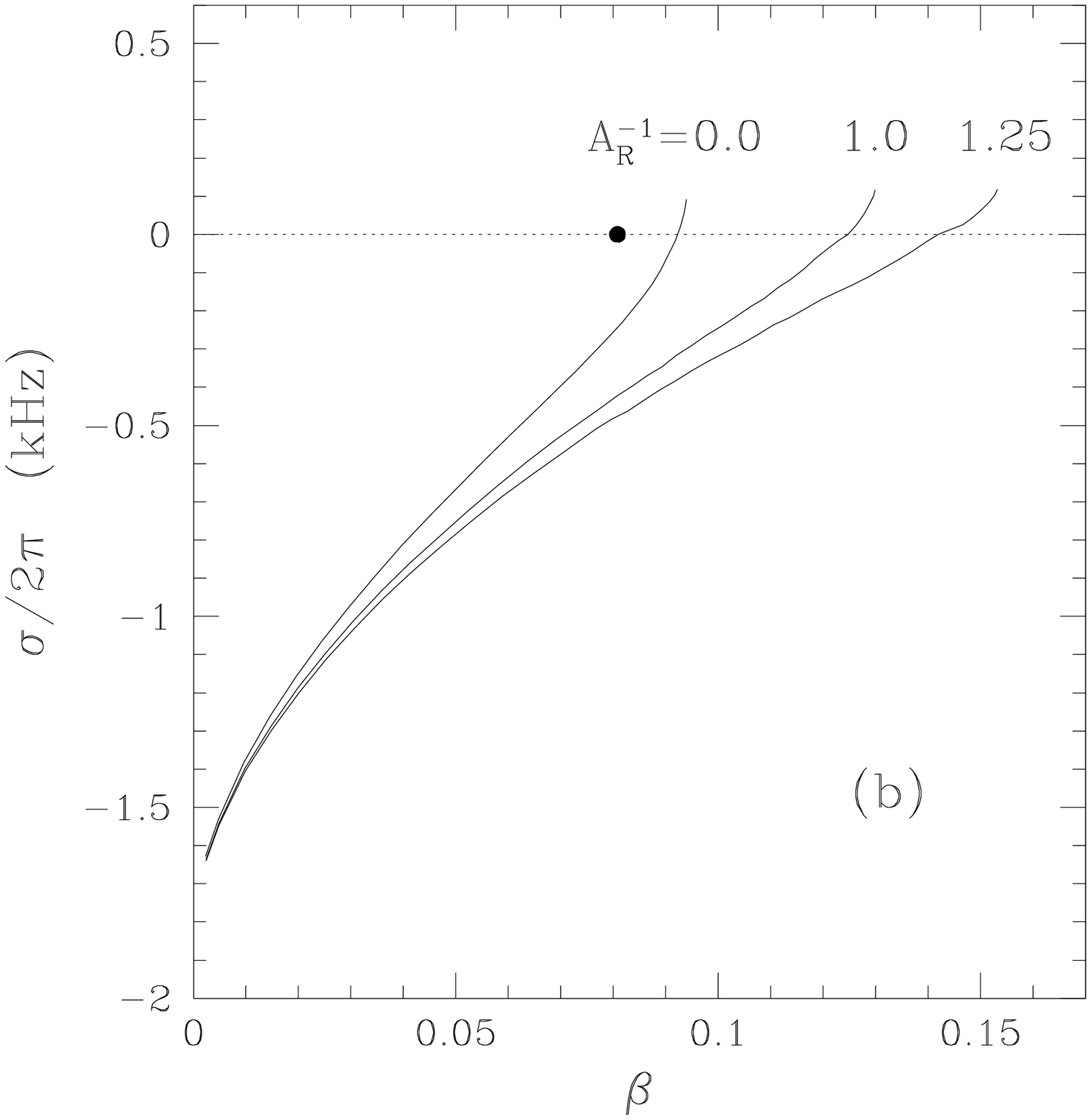}} \vskip -0.15truecm
\figcaption[]{ Eigenfrequencies of the $m=2$ mode as a
function of the parameter $\beta=T/|W|$ for the model
$(b)$ of Table~I. Different curves refer to different
degrees of differential rotation, with the $A_{_{\rm
R}}^{-1}=0.0$ line being the one of a uniformly rotating
model. The filled dot indicates the neutral stability
point of a uniformly rotating star computed in full
General Relativity (Stergioulas and Friedman 1998). }
\vskip 0.15truecm
\label{fig1}
 }

	Given the results of Fig.\ref{fig1} it is natural to ask
whether differential rotation favours or not the onset of
the $f$-mode instability in rapidly rotating relativistic
stars. The evidence that $\beta_c$ increases with
increasing differential rotation is not sufficient to
draw a conclusion since differentially rotating models
can support values of $\beta$ which are considerably
larger than those of uniformly rotating stars before
reaching the mass-shedding limit. This is simply due to
the fact that in the differentially rotating model
described by eq.~(\ref{omegagr}) the inner regions can
rotate rapidly while the outer regions rotate more
slowly, preventing mass-shedding. Thus, to quantify the
importance of differential rotation for the onset of the
instability we need to measure not only the secular
stability limit $\beta_c$, but also how close the latter
is to the ultimate limit of mass-shedding $\beta_s$. A
relative measure of the two quantities across sequences
with different degrees of differential rotation can
provide information on the likelihood of the existence of
a configuration at the onset of the secular instability
and clarify the role played by differential rotation. In
this sense, the normalization of $\beta_c$ with $\beta_s$
is equivalent to the normalization, commonly used for
uniformly rotating stars, of writing the stellar angular velocity
in terms of the mass-shedding angular velocity.

	In Fig.\ref{fig2} we show the behaviour of the ratio
${\widetilde \beta}_c\equiv\beta_c/\beta_s$ as obtained
for degrees of differential rotation ranging between
$A_{_{\rm R}}^{-1}=0$ (uniform rotation) to $A_{_{\rm
R}}^{-1}=1.25$. To avoid contamination of our results
from configurations that start showing a distinct
toroidal topology, we stop our sequences for
$A^{-1}_{_{\rm R}}=1.25$ but were able to find
eigenfrequencies also for larger values of $A^{-1}_{_{\rm
R}}$ (see Table~II).

\vspace{0.2cm}
\vbox{ \vskip 0.30truecm
\centerline{\epsfxsize=7.5truecm
\epsfbox{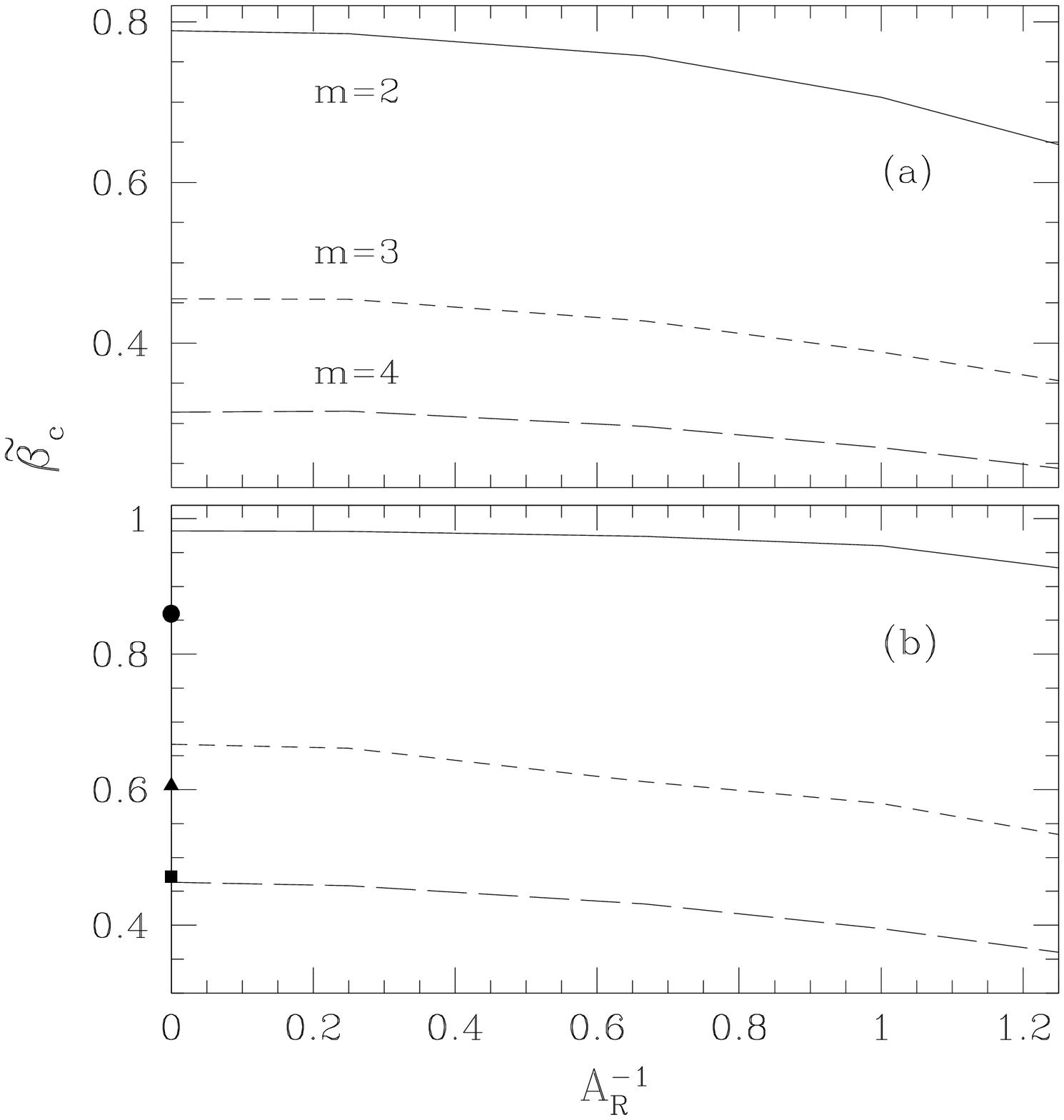}}
\vskip -0.15truecm
\figcaption[]{
${\widetilde \beta}_c$ as a function of the rate
of differential rotation. The two panels refer to the two
models of Table~I and the different line types refer to
different mode-numbers. The filled symbols show a
comparison with the fully general relativistic calculations
of Stergioulas and Friedman (1998). }
\vskip 0.15truecm
\label{fig2}
 }

	The two panels of Fig.\ref{fig2} refer to the models
$(a)$ and $(b)$ of Table~I, respectively. In each panel,
the different lines refer to the mode-numbers $m=2, 3$
and 4. As anticipated above, the Cowling approximation
overestimates the stability of the stellar models by
introducing an error which is of the order of 10\% for
the $m=2$ mode but is considerably smaller for
higher mode-numbers. This can be appreciated by comparing 
with the fully general relativistic values of
${\widetilde \beta}_c$ computed for uniformly rapidly
rotating stellar models (Stergioulas and Friedman 1998)
and indicated with filled symbols in the lower panel of
Fig.\ref{fig2}. The overall decrease of ${\widetilde \beta}_c$
for increasing degrees of differential rotation provides
a clear indication that the onset of the secular $f$-mode
instability is in general {\it favoured} by the presence
of differential rotation. This is most evident in the
case of a stiffer EOS [model $(a)$ of Table~I] where, for
reasonable degrees of differential rotation, ${\widetilde
\beta}_c$ is reduced by about 17\%. A direct comparison
with corresponding values in the literature is
presented in Table~II, where we report the values of
$\beta_c$ for different mode-numbers and degrees of
differential rotation.

	To validate our relativistic results, we have
computed the eigenfrequencies and the secular stability
limits for $f$ modes in rapidly and differentially
rotating Newtonian stellar models (A more extended
discussion of these results will be presented in
Karino et al. (2002). The numerical strategy followed is the
same as the one discussed above for the relativistic
Cowling approximation but we have considered both the
case in which perturbations in the gravitational
potential are neglected (Newtonian Cowling approximation)
and the case where they are included. The stellar models
used in this case are also polytropes with index
$N=0.5$. The results found for the corresponding
Newtonian quantity $({\widetilde \beta}_c)_{_{\rm
N}}\equiv (\beta_c/\beta_s)_{_{\rm N}}$ are presented in
Fig.\ref{fig3} where we show $({\widetilde
\beta}_c)_{_{\rm N}}$ in the Newtonian Cowling
approximation (dashed lines) and in the fully Newtonian
calculations (solid lines), for the $m=3$, $f$ mode. As
expected, the results found in Newtonian gravity are
qualitatively similar to the relativistic ones and show
that the presence of differential rotation introduces a
general decrease of $({\widetilde \beta}_c)_{_{\rm N}}$.

\vbox{
\begin{center}
\begin{tablehere}
\vspace{0.1cm}
\label{tII}
\begin{tabular}{lccc}

\hline
\hline

\medskip
$A_{_{\rm R}}^{-1}$ ~&~ $m=2$ ~&~ $m=3$  ~&~ $m=4$  \\

\hline 
     0.00  ~&~ 0.1300 ~&~ 0.0748 ~&~ 0.0516 \\
	   ~&~ 0.0923 ~&~ 0.0627 ~&~ 0.0436 \\
     0.25  ~&~ 0.1310 ~&~ 0.0761 ~&~ 0.0528 \\
	   ~&~ 0.0943 ~&~ 0.0635 ~&~ 0.0441 \\
     0.50  ~&~ 0.1410 ~&~ 0.0798 ~&~ 0.0552 \\
	   ~&~ 0.1060 ~&~ 0.0668 ~&~ 0.0471 \\
     1.00  ~&~ 0.1530 ~&~ 0.0845 ~&~ 0.0587 \\
	   ~&~ 0.1250 ~&~ 0.0753 ~&~ 0.0513 \\
     1.25  ~&~ 0.1630 ~&~ 0.0890 ~&~ 0.0615 \\
	   ~&~ 0.1420 ~&~ 0.0818 ~&~ 0.0552 \\
     1.43  ~&~ 0.1710 ~&~ 0.0927 ~&~ 0.0641 \\
	   ~&~ 0.1580 ~&~ 0.0865 ~&~ 0.0583 \\
	\hline
	\end{tabular}
\vspace{0.2cm}
\end{tablehere}
\break
\end{center}
{\small~~~~TABLE II: $\beta_c$ for different degrees of
differential rotation. The results are presented for
$m=2, 3, 4$ and the two values for each mode-number and
$A_{_{\rm R}}^{-1}$refer to models $(a)$ and $(b)$ of
Table~I, respectively.}
\vspace{0.1cm}}

	There are three aspects of Fig.\ref{fig3} that
are worth commenting on. Firstly, it should be noted that
we have not plotted the behaviour for the $m=2$ mode. 
This is
because this mode is never found unstable within the
Newtonian Cowling approximation and in the range of differential
rotation considered here. Secondly, when compared with
Fig.~\ref{fig2} the difference between the two approaches
appears to be larger for the Newtonian star. This is to
be expected since the Cowling approximation is more
accurate when the longitudinal gravitational field is
stronger and hence in General Relativity. A similar
improvement of the Cowling approximation is also observed
when the polytropic index is varied. In the case of a
softer EOS, in fact, the star is generally
more compact and the Cowling approximation more
accurate (Yoshida and Kojima 1997). Thirdly, it should be noted that
the inclusion of the perturbations in the gravitational
potential lowers systematically the values of
$({\widetilde \beta}_c)_{_{\rm N}}$ for the various
degrees of differential rotation indicating that the
Cowling approximation tends systematically to
overestimate the stability limits for the onset of the secular
$f$-mode instability. This behaviour, encountered also in
relativistic gravity, supports our expectations that a
fully general relativistic analysis, in which the metric
perturbations are not neglected, will confirm our results
and provide even more promising estimates for the secular
stability limits of $f$ modes to the CFS instability.
 
%
\vspace{0.2cm}
\vbox{ \vskip 0.30truecm
\centerline{\epsfxsize=7.5truecm
\epsfbox{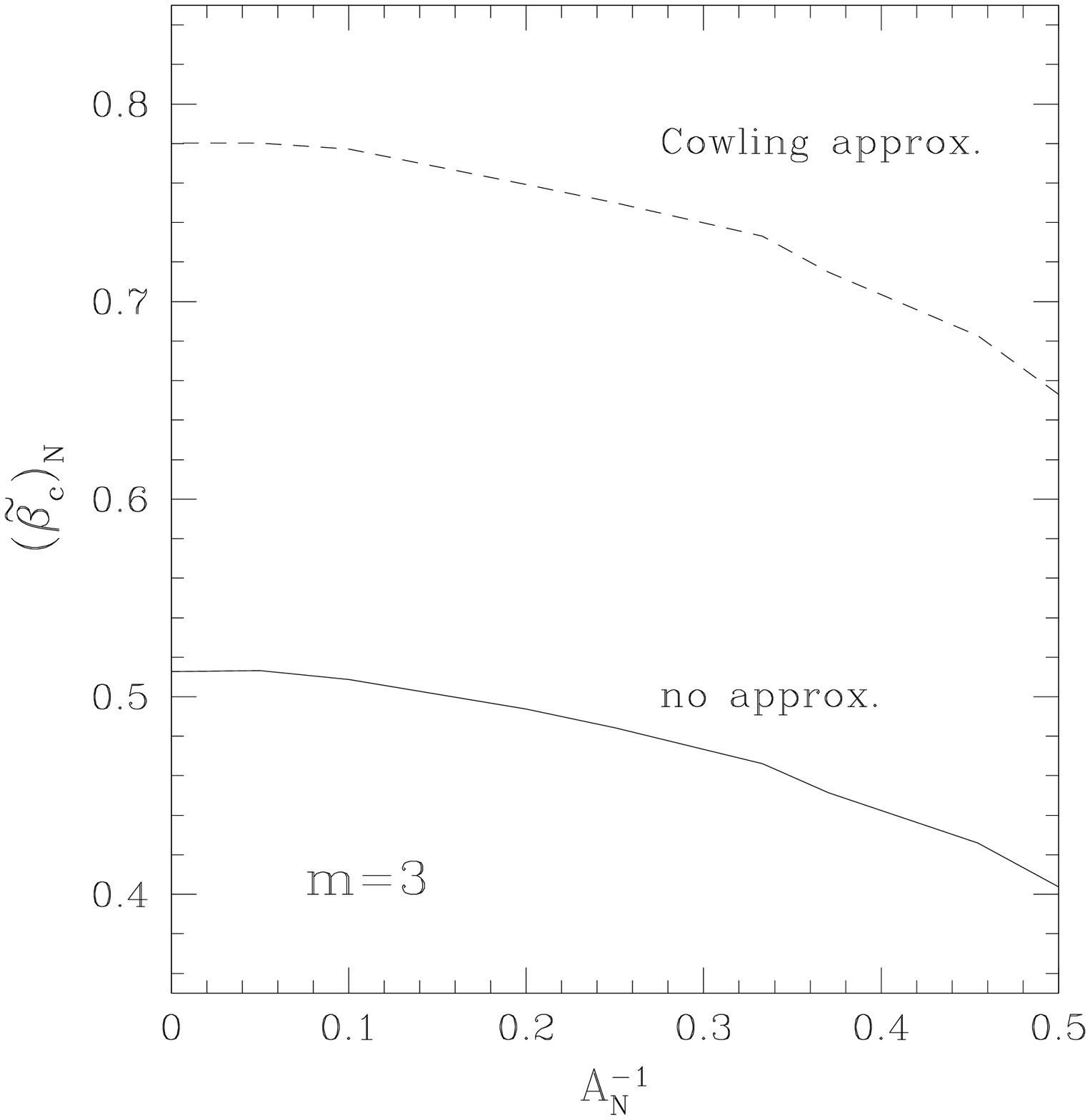}}
\vskip -0.15truecm
\figcaption[]{
$({\widetilde \beta}_c)_{_{\rm N}}$ as a function
of the Newtonian degree of differential rotation. The
different line types refer to the Newtonian Cowling
approximation (dashed lines) and to full Newtonian
calculations (solid lines).
 }
\vskip 0.15truecm
\label{fig3}
 }

\acknowledgments It's a pleasure to thank J.~Miller and
N.~Stergioulas for useful discussions and comments. This
research has been supported by the MIUR, EU Research
Training Network Contract HPRN-CT-2000-00137 and
Grant-in-Aid for Scientific Research of Japan Society for
the Promotion of Science (12640255).


\end{document}